\documentclass[aps,pra,twocolumn,superscriptaddress,10pt,article,nofootinbib,showpacs]{revtex4-1}
\usepackage{amsfonts}
\usepackage{amsmath}
\usepackage{amsthm}
\usepackage{braket}
\usepackage{graphicx}
\usepackage{amssymb}
\usepackage{dsfont}
\usepackage{mathrsfs}
\usepackage{sidecap}
\usepackage[T1]{fontenc}
\usepackage[utf8]{inputenc}
\usepackage{float}
\usepackage{eufrak}
\usepackage{leftidx}

\def \g{{\mathbf g}}
\def \h{{\mathbf h}}

\begin{document}

\title{A UV perspective on mixed anomalies at critical points between bosonic symmetry-protected phases} 

\author{Nick Bultinck}
\affiliation{Department of Physics, University of California, Berkeley, CA 94720, USA}

\begin{abstract}
Symmetry-protected phases are gapped phases of matter which are distinguished only in the presence of a global symmetry $G$. These quantum phases lack any symmetry-breaking or topological order, and have short-range entangled ground states. Based on this short-range entanglement property, we give a general argument for the existence of an emergent anti-unitary (and sometimes also a unitary) $\mathbb{Z}_2$ symmetry at a critical point separating two different bosonic symmetry-protected phases in any dimension. Often, the emergent symmetry group at criticality has a mixed global anomaly. For those phases classified by group cohomology, we identify a criterion for when such a mixed global anomaly is present, and write down representative cocycles for the corresponding anomaly class. We illustrate our results with a series of examples, and make connections to recent results on $(2+1)$-d beyond-Landau critical points.
\end{abstract}

\maketitle

\section{Introduction}

In order to understand the nature of a continuous quantum phase transition, it is crucial to identify the degrees of freedom providing the quantum fluctuations underlying the critical point \cite{Sachdev}. In the case of conventional Landau-Ginzburg phase transitions where a global symmetry gets broken spontaneously, these degrees of freedom are captured by an order parameter field, which can be used to construct a Landau free energy functional or an effective field theory. For quantum phase transitions which fall outside the Landau paradigm, identifying the degrees of freedom which take over the role of the order parameter field is one of the main challenges. 

For second order phase transitions between symmetry-protected topological (SPT) phases \cite{PollmannBerg,FidkowskiKitaev,1Dclassification,Schuch,ChenWen}, it seems especially difficult to develop a general physical picture for what is driving the critical fluctuations. This is because a SPT phase is defined to neither have symmetry breaking order, nor anyonic quasi-particles. The low-energy states of an SPT phase are therefore pretty featureless (away from the edge), so there are no distinguished bulk degrees of freedom to provide the fluctuations necessary for a continuous transition to another SPT phase. This makes the study of critical points between different SPT phases an interesting and challenging problem, which has attracted considerable attention in recent years \cite{VishwanathSenthil,ChenWang,GroverVishwanath,Lu,Tsui,Slagle,CenkeXu,YouBi,YouYou,Qin,Geraerdts,Tsui2,Verresen,You,BiSenthil,JuvenWang}.

Conceptually, one of the main insights developed in recent studies is that the idea of deconfined quantum criticality \cite{DQCP} can also be applied to transitions between SPT phases \cite{VishwanathSenthil,GroverVishwanath,Qin,Geraerdts,WangNahum,You,BiSenthil,Dualityreview}. In the deconfined quantum critical scenario, the fields capturing the critical fluctuations only accurately describe the critical point itself. The RG-relevant perturbation driving the transition confines the degrees of freedom described by these fields away from the critical point. Despite being a powerful conceptual insight, this framework does not tell us what the deconfined `fractionalized' degrees of freedom at criticality are, and so one is still forced to work on a case-by-case basis. 

The deconfined quantum criticality formalism, and other field theory studies, are expected to apply close to or at criticality, when the system has a large or infinite correlation length. In this work, we approach the problem from a different angle based on the short-range entanglement structure of gapped SPT phases. Our arguments will only hold away from the critical point, and break down once the correlation length diverges. However, as we will argue in more detail below, one can still use this approach to obtain restrictions on the field theory describing the critical point by matching the two approaches at intermediate correlation lengths.

Based on the short-range entanglement approach, we argue that at critical points between SPT phases there is always an emergent anti-unitary $\mathbb{Z}_2^{\mathcal{T}}$ symmetry, whose interplay with the microscopic global symmetry group can potentially lead to a mixed anomaly. Under a certain condition, specified below, there is also an additional emergent unitary $\mathbb{Z}_2$ symmetry at criticality, which always has a mixed anomaly with the microscopic symmetry. Because the boundaries of SPT phases are subject to global symmetry anomalies \cite{Ryu2,Wen}, this provides a general connection between continuous SPT phase transitions in $d$ dimensions and boundaries of $(d+1)$-dimensional SPTs \cite{VishwanathSenthil,Tsui}. These findings are also intimately connected to a recent study of anomalies at deconfined quantum critical points in $(2+1)$-d \cite{WangNahum} (see also \cite{Komargodski}). 

The work presented here is closely related to a previous paper by Tsui \emph{et al} \cite{Tsui}. By applying  our general formalism to the group cohomology fixed points models of Ref.~\cite{ChenWen}, one exactly reproduces the findings of Ref. \cite{Tsui}. We review this explicitly in the examples section. Also the `non-double stacking' condition identified in Ref. \cite{Tsui} plays an important role in our arguments, and its generalization is one of the main results presented here. One difference between this work and Ref. \cite{Tsui}, is that here we make use of the short-range entanglement properties of SPT phases and recent insights in the study of locality-preserving unitaries to formulate a general property of continuous phase transitions between SPT phases, which applies to both bosonic cohomology \cite{PollmannBerg,FidkowskiKitaev,1Dclassification,Schuch,ChenWen} and beyond-cohomology  \cite{BurnellChen,WangSenthil,WangSenthilU1,KapustinCobordism} SPT phases. Even restricting to phase transitions between cohomology SPT phases, our framework generalizes the findings of Ref. \cite{Tsui} to different on-site symmetry representations (see section \ref{sec:rep}). This allows us to study more general examples, and connect to topics discussed in recent studies of $(2+1)$-d beyond-Landau critical points, such as dualities \cite{WangSenthilX,MetlitskiVishwanath,Karch,Mross,XuYou,Hsin,Seiberg,WangNahum,Dualityreview} and symmetric mass generation \cite{BenTov,Catterall,Venkitesh,Schaich,Ayyar,HeWu,Huffman,YouHeXu,You}.

Before turning to the main discussion, let us first introduce some terminology and recall a couple of facts about SPT phases. We will refer to a SPT phase protected by a global symmetry group $G$ as a $G$-SPT phase. We only consider those $G$-SPT phases that can be trivialized by explicitly breaking $G$, which in particular means that all $G$-SPT phases are non-chiral. The $G$-SPT phases form an abelian group, where the group multiplication is the stacking operation. On the level of ground states, the stacking of two SPT ground states $|\psi_1\rangle$ and $|\psi_2\rangle$ simply refers to taking the tensor product $|\psi_1\rangle\otimes|\psi_2\rangle$. Importantly, the group structure also means that $G$-SPT phases are invertible, i.e. for every $G$-SPT phase there exists an inverse $G$-SPT phase such that under stacking they combine to the trivial phase. This invertibility is closely related to the entanglement structure of SPT ground states, as short-range entangled ground states are ground states of invertible phases \cite{KitaevPreskill,LevinWen,ChenGu}. The group structure of $G$-SPT phases significantly simplifies the study of phase transitions between $G$-SPT phases, as one can without loss of generality restrict to phase transitions between the trivial phase and a non-trivial SPT phase. This is because one can always stack the system of interest with a $G$-SPT phase to make one side of the transition trivial, without changing the quantum critical fluctuations. We will come back to this point at the end of the manuscript. But for now, let us focus only on phase transitions between the trivial phase and a non-trivial $G$-SPT.

The results of this manuscript are organized below in the following way. In Section \ref{squaringTriv}, we first consider critical points between the trivial phase a non-trivial $G$-SPT phase that squares to the trivial phase. It is argued that such critical points come with an emergent unitary $\mathbb{Z}_2$ symmetry, which always has a mixed anomaly with the microscopic global symmetry group $G$. Section \ref{antiUnitary} deals with transitions between the trivial phase and a general $G$-SPT phase. In this general case, we find that if the transition is continuous, then there is an emergent anti-unitary $\mathbb{Z}_2^{\mathcal{T}}$ symmetry at criticality. Depending on the type of local symmetry representation of $G$, the minimal emergent symmetry group is either $\mathbb{Z}_2^{\mathcal{T}}\times G$ or $\mathbb{Z}_2^{\mathcal{T}}\rtimes G$. The anti-unitary symmetry $\mathbb{Z}_2^{\mathcal{T}}$ can have a mixed anomaly with the microscopic symmetry group $G$, but this is not always the case. For both possible emergent symmetry groups ($\mathbb{Z}_2^{\mathcal{T}}\times G$ or $\mathbb{Z}_2^{\mathcal{T}}\rtimes G$), we give a necessary and sufficient criterion for a non-trivial mixed anomaly between $\mathbb{Z}_2^{\mathcal{T}}$ and $G$ to be present (provided that the non-trivial SPT phase involved is classified by group cohomology \cite{ChenWen}). Note that by combining the results of Sections \ref{squaringTriv} and \ref{antiUnitary}, we arrive at the conclusion that if the non-trivial SPT phase squares to the trivial phase, then the critical point will have an emergent $\mathbb{Z}_2\times\mathbb{Z}_2^{\mathcal{T}}$ symmetry. Section \ref{examples} contains a collection of concrete models such as spin chains, cohomology fixed-point models and coupled-wire constructions that serve as examples for the general theory. We end with a discussion of the results and open questions in Section \ref{discussion}.

\section{SPT phases squaring to the trivial phase}\label{squaringTriv}

\begin{center}
\begin{figure}
\includegraphics[scale=0.6]{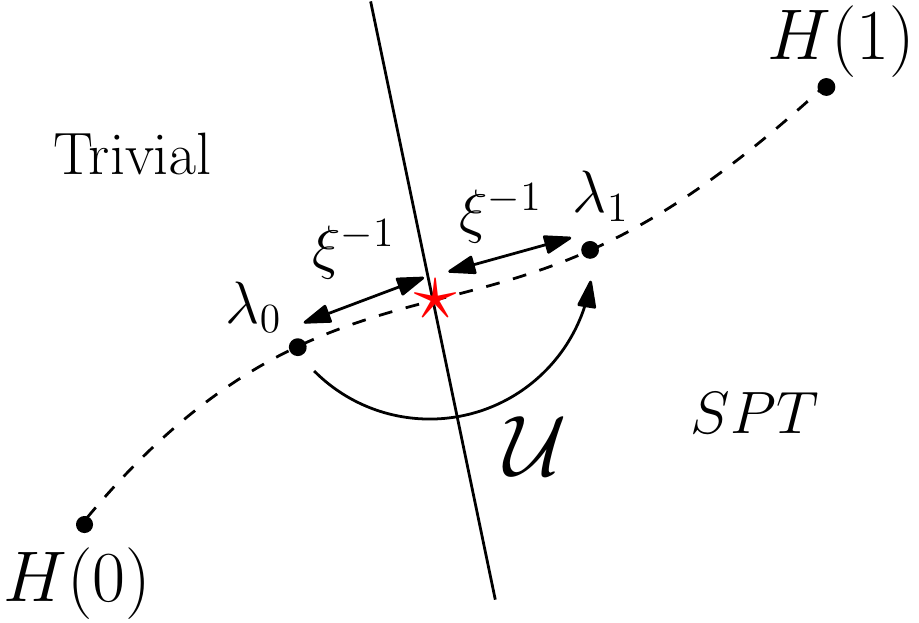}
\caption{A path of local $G$-symmetric, short-range entangled bosonic Hamiltonians $H(\lambda)$, with a critical point at $\lambda=\lambda^*$, depicted by the red star. For $\lambda<\lambda^*$ ($\lambda>\lambda^*$), the Hamiltonian realizes a trivial (non-trivial) $G$-SPT phase. For two Hamiltonians $H(\lambda_0)$ and $H(\lambda_1)$ at the same distance $\xi^{-1}_0=\xi^{-1}_1$ from the critical point, there exists a locality preserving unitary $\mathcal{U}$ mapping the ground state of $H(\lambda_0)$ to the ground state of $H(\lambda_1)$.}\label{fig:path}
\end{figure}
\end{center}

We consider a one-parameter family of local spin or boson lattice Hamiltonians $H(\lambda)$ $(\lambda\in[0,1])$ defined with periodic boundary conditions. For each value of $\lambda$, the Hamiltonian $H(\lambda)$ has a global symmetry $G$, i.e. $[H(\lambda),U(\textbf{g})^{\otimes N}]=0$ for all $\lambda$ and $\g\in G$. Here, $U(\g)$ is the local unitary symmetry action on each site, and $N$ is the number of sites in the system. We choose the one-parameter family of Hamiltonians such that $H(0)$ realizes a trivial symmetric quantum phase, meaning that its unique ground state is continuously connected to a product state by a path of symmetric short-range entangled states. $H(1)$ realizes a non-trivial short-range entangled quantum phase protected by the global symmetry $G$. This means that $H(1)$ realizes a trivial quantum phase once the global symmetry is allowed to be broken. We further assume that the non-trivial $G$-SPT phase is separated from the trivial phase by a single critical point at $\lambda=\lambda^*$. We will denote the `distance' of a Hamiltonian $H(\lambda)$ to the critical point at $\lambda^*$ by its inverse zero-temperature correlation length $\xi_\lambda^{-1}$. See Fig. \ref{fig:path} for an illustration of the Hamiltonian path $H(\lambda)$.

Now consider two Hamiltonians $H(\lambda_0)$ and $H(\lambda_1)$ at the same distance from the critical point, such that $\lambda_0<\lambda^*<\lambda_1$. Because the unique ground states of both $H(\lambda_0)$ and $H(\lambda_1)$ are short-range entangled \cite{ChenGu}, it is believed, based on many examples, that there exists a locality-preserving unitary $\mathcal{U}$ that maps the ground state of $H(\lambda_0)$ to the ground state of $H(\lambda_1)$ (up to arbitrarily small error). A locality-preserving unitary is a unitary operator satisfying $\mathcal{U}^\dagger \mathcal{O}_{\mathcal{R}} \mathcal{U} \subset \mathcal{O}_{\mathcal{R}'}$, where $\mathcal{O}_{\mathcal{R}}$ is the operator algebra supported on the sites in a finite region $\mathcal{R}$, and $\mathcal{R}\subset \mathcal{R}'$. A locality-preserving unitary preserves the correlation length, as can easily be seen from the following equality:

\begin{equation}
\langle \mathcal{U}^\dagger O_{\mathcal{R}_m} O_{\mathcal{R}_n}\mathcal{U}\rangle = \langle O_{\mathcal{R}'_m} O_{\mathcal{R}'_n}\rangle\, ,
\end{equation}
where $\mathcal{R}_n$ denotes the region $\mathcal{R}$ centered at the site labeled by $n$, and $O_{\mathcal{R}_n}$ is an operator with support on the sites in $\mathcal{R}_n$. The existence of such a locality-preserving unitary that maps short-range entangled states `across a critical point' is a novel feature of phase transitions between SPT phases and is not shared by conventional Landau-Ginzburg symmetry-breaking phase transitions. Indeed, even for large but finite system sizes where a spontaneous symmetry breaking quantum phase has a unique symmetric ground state (seperated by an energy gap decreasing in the system size), there cannot exist a locality-preserving unitary mapping the ground state of the symmetric phase to the ground state of the symmetry-broken phase, as the latter has long-range order \cite{Verstraete}. 

The locality-preserving unitary $\mathcal{U}$ by definition has the property that when acting on a trivial symmetric product state, it transforms it into a non-trivial SPT ground state. Locality-preserving unitaries with this property have recently been studied extensively in the context of Floquet systems, where they have the physical intepretation of pumping a lower-dimensional SPT phase to the boundary of the sample during a single drive period \cite{Else2,Sondhi,Gross,PoFidkowskiMorimoto,FidkowskiPo,PoFidkowski,Cirac,Sahinoglu}. Based on these studies, we conjecture that there will always exist a locality-preserving unitary $\mathcal{U}$ mapping the ground state of $H(\lambda_0)$ to the ground state of $H(\lambda_1)$ that commutes with the global symmetry $G$. 

Up to now, we have argued that there exists a locality-preserving unitary $\mathcal{U}$ satisfying
\begin{equation}
\mathcal{U}|\psi_0\rangle = |\psi_1\rangle\,,\hspace{0.3 cm} [\mathcal{U},U(\g)^{\otimes N}] = 0\, , \forall \g\, ,
\end{equation}
where $|\psi_0\rangle$ ($|\psi_1\rangle)$ is the unique symmetric ground state of $H(\lambda_0)$ ($H(\lambda_1)$). At this point we restrict to the special case where the non-trivial SPT phase realized by $H(\lambda_1)$, `squares' to the trivial phase, with which we refer to the group structure of $G$-SPT phases. We thus, in this section, restrict to SPT phases with the property that if we stack two copies of the system, the resulting system forms a trivial SPT phase. This restriction implies that if we apply $\mathcal{U}$ twice on $|\psi_0\rangle$, the resulting state is also a trivial SPT ground state with the same correlation length. This follows because the $G$-SPT invariants associated with $\mathcal{U}$ are multiplicative under multiplication of $\mathcal{U}$ \cite{Else2,Sondhi,Gross,PoFidkowskiMorimoto,FidkowskiPo,PoFidkowski}. In general, $\mathcal{U}^2|\psi_0\rangle$ will nevertheless be different from $|\psi_0\rangle$. However, $|\psi_0\rangle$ and $\mathcal{U}^2|\psi_0\rangle$ are different only in their short-distance properties and this difference will be washed out by any renormalization-group machinery we use to arrive at a CFT describing the universality class of the critical point. Because $\mathcal{U}$ commutes with $G$ and therefore preserves symmetry quantum numbers, we can also use $\mathcal{U}$ to map the low-energy quasi-particle states of $H(\lambda_0)$ to those of $H(\lambda_1)$ (again up to short-distance properties). From the perspective of the CFT at $\lambda^*$, $\mathcal{U}$ therefore acts as a $\mathbb{Z}_2$ symmetry, resulting in a minimal symmetry group $G\times\mathbb{Z}_2$ at the critical point. The trivial and non-trivial SPT phases are obtained by turning on a relevant perturbation that \emph{explicitly} breaks this $\mathbb{Z}_2$ symmetry. We emphasize that in the RG fixed point theory $\mathcal{U}$ acts as a symmetry and not as a duality, as a duality does not map local operators to local operators.

An important property of the symmetry group $G\times \mathbb{Z}_2$ at the critical point is that it is realized in an intrinsically non-local way, as we now explain. Of course, $G$ by itself is implemented locally via the on-site symmetry action $U(\g)$. However, because $H(\lambda_1)$ realizes a non-trivial SPT phase, we know that $[\mathcal{U},U(\g)^{\otimes N}]=0$ cannot be true with open boundaries, because of the non-trivial edge physics of SPT phases. To see this, note that the trivial ground state has both short-range correlations and short-range entanglement along the edge or boundary, while the non-trivial SPT ground state either has off-diagonal long-range order, algebraic decay of correlations or anyonic excitations on a symmetric boundary. So we conclude that the commutator $[\mathcal{U},U(\g)^{\otimes N}]=0$ cannot be realized locally. There are two main possibilities for why the commutator $[\mathcal{U},U(\g)^{\otimes N}]$ is non-zero with open boundary conditions. In the first possibility, $\mathcal{U}$ is a finite-depth quantum circuit, meaning that it is of the form

\begin{equation}
\mathcal{U}=\prod_{\alpha=1}^D U_{\alpha}\, , 
\end{equation}
where each $U_{\alpha}$ is a layer of non-overlapping unitaries with strictly local support, called `gates', and $D$ is a system-size independent number called the `depth' of the circuit. The non-locality comes from the fact that although $\mathcal{U}$ commutes with $G$, it is impossible to find individual gates that commute with $G$. So in this case, it is possible to restrict $\mathcal{U}$ to a finite region $\mathcal{R}$ while preserving unitarity by simply keeping only those gates that lie entirely within $\mathcal{R}$, but the restricted $\mathcal{U}$ no longer commutes with the global symmetry. This scenario is believed to apply to symmetry-protected phases classified by group cohomology \cite{ChenWen,ChenGu}. The second possibility is that $\mathcal{U}$ is non-trivial as a locality-preserving unitary, meaning that it cannot be represented by a finite-depth quantum circuit. In Ref. \cite{Haah} the authors presented strong arguments that this second scenario applies to beyond-cohomology \cite{BurnellChen,WangSenthil,WangSenthilU1,KapustinCobordism} SPT phases. In this work, we will restrict ourselves to transitions between group cohomology SPT phases and defer the beyond-cohomology phases to future studies. 

Because the symmetry group $G\times \mathbb{Z}_2$ is realized in an intrinsically non-local way on the lattice, it can potentially have a mixed anomaly, meaning that no short-range entangled state can be invariant under the $G\times\mathbb{Z}_2$ symmetry action. In fact, we expect this to be the case almost by definition because any $G$-symmetric short-range entangled state is the ground state of a particular $G$-SPT phase, while $\mathcal{U}$ maps different $G$-SPT phases to each other. 

In order to make the identification of the mixed anomaly more precise, we recall that $\mathcal{H}^{d+2}[G\times \mathbb{Z}_2,U(1)]$, the $(d+2)$-th Borel cohomology group with $U(1)$ coefficients, labels global symmetry anomalies in $d$ spatial dimensions \cite{DijkgraafWitten,ChenWen,Wen}. Using the K\"unneth formula \cite{WenK}, we can write the cohomology group of a product group $G\times H$ as

\begin{equation}
\mathcal{H}^n[G\times H,U(1)] =\bigoplus_{k=0}^n \mathcal{H}^k[H,\mathcal{H}^{d-k}[G,U(1)]]
\end{equation}
The mixed anomaly of the symmetry group $G\times \mathbb{Z}_2$ at the critical point corresponds to a non-trivial element of

\begin{equation}\label{H1}
\mathcal{H}^1[\mathbb{Z}_2,\mathcal{H}^{d+1}[G,U(1)]]\, ,
\end{equation}
which labels the homomorphisms from $\mathbb{Z}_2$ to $\mathcal{H}^{d+1}[G,U(1)]$. Because we assumed that the SPT phase under consideration squares to the trivial phase, $\mathcal{H}^{d+1}[G,U(1)]$ has a $\mathbb{Z}_2$ subgroup, which is the image of the homomorphism corresponding to the non-trivial class in $\mathcal{H}^1[\mathbb{Z}_2,\mathcal{H}^{d+1}[G,U(1)]]$ responsible for the mixed anomaly. Note that Eq. \eqref{H1} is also the element of $\mathcal{H}^{d+2}[G\times \mathbb{Z}_2, U(1)]$ classifying $(d+1)$-dimensional cohomology SPT phases that can be obtained via the decorated domain wall construction \cite{ChenLu}. When locality-preserving unitaries are interpreted as `SPT pumps' in Floquet systems, the class in $\mathcal{H}^{d+1}[G,U(1)]$ that is the image of the non-trivial group element of $\mathbb{Z}_2$ under the above homomorphism labels the SPT that gets pumped to the boundary during the drive period \cite{Else2,Sondhi,PoFidkowskiMorimoto,FidkowskiPo,PoFidkowski}.

In Ref. \cite{ChengZaletel}, the following explicit representation for the $n$-cocycles of $G\times H$ was given:

\begin{align}\label{repr}
\omega_n((\g_1,\h_1),\dots,(\g_n,\h_n))= \nonumber \\  \prod_{k=0}^n \nu_k(\h_1,\dots,\h_k;\g_{k+1},\dots,\g_n)\,,
\end{align}
where $(\g,\h)$ with $\g\in G$, $\h\in H$ represents a general group element of $G\times H$, and each $\nu_k$ satisfies the $n$-cocycle relation $\mathrm{d}\nu_k=1$ (see Appendix \ref{AppCoh} for the definition of $\mathrm{d}$). For $H=\mathbb{Z}_2$, the $(d+2)$-cocycle that is a representative of the mixed anomaly class in $\mathcal{H}^1[\mathbb{Z}_2,\mathcal{H}^{d+1}[G,U(1)]]$ is given by $\nu_1$, for which the $(d+2)$-cocycle relation takes the following form \cite{ChengZaletel}:

\begin{align}\label{cocycle}
\nu_1(\h_2;\g_3,\dots,\g_{d+3})\nu_1^{(-1)^{d+3}}(\h_1;\g_2,\dots,\g_{d+2}) \nonumber \\
\times \prod_{q=2}^{d+2}\nu_1^{(-1)^q}(\h_1;\g_2,\dots,\g_{q}\g_{q+1},\dots,\g_{d+3})\nonumber \\
\times \nu_1^{-1}(\h_1\h_2;\g_3,\dots,\g_{d+3})=1\, 
\end{align}
This expression allows us to obtain a representative $(d+2)$-cocycle for the mixed anomaly class in $\mathcal{H}^1[\mathbb{Z}_2,\mathcal{H}^{d+1}[G,U(1)]]$, given a representative $(d+1)$-cocycle of the class in $\mathcal{H}^{d+1}[G,U(1)]$ corresponding to the non-trivial SPT phase. Indeed, if we take $\nu_1(\h_1;\g_2\,\dots,\g_{d+2})$ with $\h_1$ fixed to be a $(d+1)$-cocycle of G, then Eq. \eqref{cocycle} reduces to

\begin{align}\label{homom}
\nu_1(\h_2;\g_3,\dots,\g_{d+2})\nu_1(\h_1;\g_3\dots,\g_{d+2}) \nonumber \\
\times \nu_1^{-1}(\h_1\h_2;\g_3,\dots,\g_{d+2})=1
\end{align}
If a cohomology class $[\omega]\in\mathcal{H}^{d+1}[G,U(1)]$ squares to the trivial class, then every representative $(d+1)$-cocycle $\omega_{d+1} \in [\omega]$ satisfies $\omega_{d+1}^2=\mathrm{d}\beta_{d}$, where $\beta_d$ is a $d$-cochain (see Appendix \ref{AppCoh} for a detailed review of group cohomology). There always exists a coboundary transformation $\omega_{d+1}\rightarrow \omega'_{d+1}=\omega_{d+1}\mathrm{d}\beta_{d}^{-1/2}$ such that $\left(\omega'_{d+1}\right)^2\equiv 1$. So given a representative $(d+1)$-cocyle $\omega_{d+1}$ of any cohomology class $[\omega]\in\mathcal{H}^{d+1}[G,U(1)]$ squaring to the trivial class, we can always construct a $(d+2)$-cocycle of $G\times\mathbb{Z}_2$ by applying the coboundary transformation $\omega_{d+1}\rightarrow \omega'_{d+1}$, and defining $\nu_1(\h,\g_1,\dots,\g_{d+2})\equiv \omega'_{d+1}(\g_1,\dots,\g_{d+2})$ if $\h$ is the non-trivial group element of $\mathbb{Z}_2$ and $\nu_1(\h,\g_1,\dots,\g_{d+2})\equiv 1$ otherwise. With this definition, one can check that Eq. \eqref{homom} is satisfied. In the supplementary material we explicitly verify for $d=1$ that $\nu_1$ defined in this way is indeed a representative cocycle for the cohomology class associated with the mixed $G\times\mathbb{Z}_2$ anomaly at the critical point, using a generalization of the dimensional reduction procedure developed in Ref. \cite{Else}.

\section{Transitions to general SPT phases} \label{antiUnitary}

In the previous section, we crucially relied on the assumption that a certain non-trivial SPT phase squares to the trivial phase in order to arrive at the conclusion that there is an emergent $G\times \mathbb{Z}_2$ symmetry at a critical point separating this SPT phase from the trivial phase. In this section, we consider the cases where the non-trivial SPT does not square to the trivial phase. 

We again consider a one-parameter family of local Hamiltonians $H(\lambda)$ ($\lambda\in[0,1]$) with global symmetry $G$, where $H(0)$ realizes a trivial SPT phase and $H(1)$ a non-trivial SPT phase, which now does not square to the trivial phase. As before, we assume that both quantum phases are separated by a single critical point at $\lambda=\lambda^*$. For bosonic SPT phases, complex conjugation `inverts' the quantum phase, meaning that the local spin or boson Hamiltonians $H$ and $H^*$ realize phases which are each others inverse under the stacking operation (if they have short-range entangled ground states). This motivates us to consider locality-preserving unitaries $\mathcal{U}$ commuting with $U(\g)^{\otimes N}$ that would map the \emph{complex conjugate} ground state of $H(\lambda_0)$ to the ground state of $H(\lambda_1)$, where as before $H(\lambda_0)$ and $H(\lambda_1)$ are two Hamiltonians at the same distance from the critical point with $\lambda_0<\lambda^*<\lambda_1$. Let us again denote the ground state of $H(\lambda_0)$ ($H(\lambda_1)$) as $|\psi_0\rangle$ ($|\psi_1\rangle$). The reason for including the complex conjugation is that iterating this operation twice on $|\psi_0\rangle$ gives $\mathcal{U}\mathcal{U}^*|\psi_0\rangle$, which is also a trivial SPT ground state. So using the same logic as in the previous section, we would now like to conclude that at the critical point there is an additional \emph{anti-unitary} $\mathbb{Z}_2$ symmetry. However, in order to make this statement there remains one important point we need to take into account. In general, it is not guaranteed that $|\psi_0\rangle^*$ is invariant under $U(\g)^{\otimes N}$. If $|\psi_0\rangle^*$ is not invariant under $U(\g)^{\otimes N}$, the equality $|\psi_1\rangle=\mathcal{U}|\psi_0\rangle^*$ obviously cannot be true if $\mathcal{U}$ commutes with the global symmetry action. In order to address this issue, additional information about the type of local symmetry representation $U(\g)$ is required. Below, we consider two different scenarios depending on the properties of $U(\g)$, and discuss how the symmetry group at criticality differs in both cases.

\subsection{Real and pseudo-real symmetry representations}

Let us here assume that the local symmetry representation $U(\g)$ is either real or pseudo-real, implying that there exists a unitary matrix $T$ such that:

\begin{equation}
U(\g)^*=T^\dagger U(\g)T\, ,
\end{equation}
Using $T$, let us now define the following anti-unitary locality-preserving operator

\begin{equation}
\hat{\mathcal{U}}=\mathcal{U}T^{\otimes N}K\, ,
\end{equation}
where $K$ denotes complex conjugation, and, as before, $\mathcal{U}$ is a locality-preserving unitary commuting with $G$, which we choose such that $\hat{\mathcal{U}}|\psi_0\rangle = |\psi_1\rangle$. By construction, $\hat{\mathcal{U}}$ satisfies

\begin{equation}
[\hat{\mathcal{U}},U(\g)^{\otimes N}]=0\, ,
\end{equation}
and $\hat{\mathcal{U}}^2|\psi_0\rangle$ is a trivial SPT ground state. From this we conclude that the CFT describing the critical point will have a $G\times\mathbb{Z}_2^\mathcal{T}$ symmetry, where we use the notation $\mathbb{Z}_2^\mathcal{T}$ to denote a $\mathbb{Z}_2$ group where the non-trivial group element is anti-unitary. 

In the previous section, we argued that when the non-trivial SPT phase squares to the trivial phase, the $G\times\mathbb{Z}_2$ symmetry at the critical point always has a mixed anomaly. In the present case, with symmetry group $G\times \mathbb{Z}_2^\mathcal{T}$ at criticality, the same can happen because $[\hat{\mathcal{U}},U(\g)^{\otimes N}]$ is again realized in an intrinsically non-local way.  However, now the action of $\hat{\mathcal{U}}$ on the group of $G$-SPT phases is different as compared to the previous section: it first maps a $G$-SPT phase to its inverse via the complex conjugation, and then multiplies it with the non-trivial SPT phase realized by $H(\lambda_1)$. In this case it is possible for this action to have a short-range entangled $G$-SPT fixed point. If we denote the $G$-SPT phase realized by $H(\lambda_1)$ by its cohomology class $[\omega]$, then in order for a $G$-SPT phase $[\tilde{\omega}]$ to be a fixed point of $\hat{\mathcal{U}}$, it should satisfy $[\omega]\cdot[\tilde{\omega}]^{-1}=[\tilde{\omega}]$, which is possible if $[\omega]=[\tilde{\omega}]^2$. This is the `(non-) double stacking condition' identified previously in Ref. \cite{Tsui}. So when $[\omega]=[\tilde{\omega}]^2$, we do not expect the symmetry group $G\times \mathbb{Z}_2^\mathcal{T}$ to have a mixed anomaly, as there is a potential short-range entangled $G$-SPT phase that can be invariant under it.

The relevant $G\times\mathbb{Z}_2^\mathcal{T}$ anomalies in $d$ spatial dimensions correspond to elements of the cohomology group $\mathcal{H}^{d+2}[G\times\mathbb{Z}_2^\mathcal{T},U(1)^*]$, where U$(1)^*$ means that $\mathbb{Z}_2^\mathcal{T}$ acts non-trivially on the U$(1)$ coefficients of the cohomology group \cite{ChenWen}. Because of this non-trivial action on the U$(1)$ coefficients, we cannot use the K\"unneth formula to decompose $\mathcal{H}^{d+2}[G\times\mathbb{Z}_2^\mathcal{T},U(1)^*]$. However, we can still use the explicit representation for the $n$-cocycles of $G\times\mathbb{Z}_2^\mathcal{T}$ given in Ref. \cite{ChengZaletel}. In particular, we can again write a $G\times\mathbb{Z}_2^\mathcal{T}$ $n$-cocycle $\omega_n$ as a product of $\nu_k$ as in Eq. \eqref{repr}, where each $\nu_k$ now satisfies the $n$-cocycle relation with the appropriate action on the U$(1)$ coefficients. As before, the $(d+2)$-cocycle corresponding to the possible mixed anomaly at the critical point corresponds to $\nu_1$, for which the $(d+2)$-cocycle relation now takes the form

\begin{align}\label{twcoc}
\nu_1^{\gamma(\h_1)}(\h_2;\g_3,\dots,\g_{d+3})\nu_1^{(-1)^{d+3}}(\h_1;\g_2,\dots,\g_{d+2}) \nonumber \\
\times \prod_{q=2}^{d+2}\nu_1^{(-1)^q}(\h_1;\g_2,\dots,\g_{q}\g_{q+1},\dots,\g_{d+3})\nonumber \\
\times \nu_1^{-1}(\h_1\h_2;\g_3,\dots,\g_{d+3})=1
\end{align}
Here we have introduced the following notation to denote whether a group element is unitary or anti-unitary:
\begin{equation}
\gamma(\h)=\begin{cases} 1, & \h \text{ is unitary} \\ -1, & \h \text{ is anti-unitary}  \end{cases}
\end{equation}
In the present context, $H=\mathbb{Z}_2^\mathcal{T}$, such that $\gamma=-1$ for the non-trivial group element, and $\gamma=1$ for the identity group element. The advantage of the explicit form for $\nu_1$ is again that it allows us to construct a $(d+2)$-cocycle of $G\times \mathbb{Z}_2^\mathcal{T}$ given a $(d+1)$-cocycle of $G$. Indeed, if we define $\nu_1$ similarly as before as

\begin{equation}\label{constr}
\nu_1(\h_1;\g_2,\dots,\g_{d+2})=\omega_{d+1}^{(1-\gamma(\h_1))/2}(\g_2,\dots,\g_{d+2})\, ,
\end{equation}
where $\omega_{d+1}$ is any $(d+1)$-cocycle of $G$, then we see that the $(d+2)$-cocycle relation in Eq. \eqref{twcoc} is identically satisfied. Based on the arguments above, we now expect that $\nu_1$ defined in Eq. \eqref{constr} is trivial when $\omega_{d+1}=\tilde{\omega}^2_{d+1}$, where $\tilde{\omega}_{d+1}$ is itself a $(d+1)$-cocycle. Indeed, let us in this case define

\begin{equation}
\beta_{d+1}((\g_1,\h_1),\dots,(\g_{d+1},\h_{d+1}))= \tilde{\omega}_{d+1}^{-1}(\g_1,\dots,\g_{d+1})
\end{equation}
One can check that with this definition, $\nu_1$ can be written as

\begin{align}\label{triv}
\nu_1(\h_1;\g_2,\dots,\g_{d+2}) = \nonumber \\ \beta_{d+1}^{\gamma(\h_1)}((\g_2,\h_2),\dots,(\g_{d+2},\h_{d+2})) 
\nonumber \\\times \beta^{(-1)^{d+2}}_{d+1}((\g_1,\h_1),\dots,(\g_{d+1},\h_{d+1}))\nonumber\\
\prod_{q=2}^{d+2}\beta_{d+1}^{(-1)^{q-1}}(\dots,(\g_{q-1}\g_q,\h_{q-1}\h_q),\dots,(\g_{d+2},\h_{d+2}))
\end{align}
implying it is a trivial $(d+2)$-cocycle of $G\times\mathbb{Z}_2^\mathcal{T}$. Note in particular that this implies that $\nu_1$ is a trivial $(d+2)$-cocycle if $\omega_{d+1}$ is a trivial $(d+1)$-cocycle, as in that case one can always take the square root of $\omega_{d+1}$ and still satisfy the $(d+1)$-cocycle relation. It is not hard to see that the opposite implication is also true, i.e. if $\nu_1$ is trivial, then $\omega_{d+1}=\tilde{\omega}_{d+1}^2$ for some $(d+1)$-cocycle $\tilde{\omega}_{d+1}$. To arrive at this conclusion, assume that Eq. \eqref{triv} holds, with $\nu_1$ defined as in Eq. \eqref{constr}. This immediately implies that $\beta$ can only be a function of the group elements of $G$. By taking $\h_1$ to be the trivial group element such that $\gamma(\h_1)=1$, one finds that $\beta$ is a $(d+1)$-cocycle of $G$. By taking $\h_1$ to be the non-trivial group element of $\mathbb{Z}_2^\mathcal{T}$ it then follows that $\omega=\beta^{-2}$.

\subsection{Complex symmetry representations}\label{sec:rep}

In this section we consider the case where the on-site symmetry representation of $G$ is neither real nor pseudo-real. We instead assume that the local symmetry representation of $G$ satisfies

\begin{equation}\label{auto}
U(\g)^*=T^\dagger U(\pi(\g))T\, ,
\end{equation}
where $T$ is again a unitary matrix and $\pi$ is an automorphism of $G$. As before, we define $\hat{\mathcal{U}}=\mathcal{U}T^{\otimes N}K$, with $\mathcal{U}$ a locality-preserving unitary commuting with $U(\g)^{\otimes N}$, such that $\hat{\mathcal{U}}|\psi_0\rangle = |\psi_1\rangle$ and $\hat{\mathcal{U}}^2|\psi_0\rangle$ is a trivial SPT ground state. The difference with the previous section is that now $\hat{\mathcal{U}}$ does not commute with $U(\g)^{\otimes N}$, but it instead implements the automorphism $\pi$.

Let us define $ \h \mapsto \pi_\h$ to be a homomorphism from $H$ to the automorphism group of $G$, meaning that it satisfies

\begin{eqnarray}
\pi_\h(\g_1)\pi_\h(\g_2) & = & \pi_\h(\g_1\g_2)\\
 \pi_{\h_1}(\g)\pi_{\h_2}(\g)& = & \pi_{\h_1\h_2}(\g)
\end{eqnarray}
We will consider the case where $H=\mathbb{Z}_2$, such that the automorphism corresponding to the non-trivial group element of $H$ is given by $\pi$ as in Eq. \eqref{auto}. Because of the non-trivial automorphism induced by $\hat{\mathcal{U}}$, the symmetry group at the critical point will be $G\rtimes \mathbb{Z}_2^\mathcal{T}$, where the group elements, denoted as $(\g,\h)$, satisfy the multiplication rule

\begin{equation}
(\g_1,\h_1)(\g_2,\h_2)=(\g_1\pi_{\h_1}(\g_2),\h_1\h_2)
\end{equation}

Also with a non-trivial homomorphism $\pi$, $\hat{\mathcal{U}}$ still acts on the group of $G$-SPT phases as $[\tilde{\omega}]\rightarrow [\omega]\cdot[\tilde{\omega}]^{-1}$, with $[\omega]$ the $G$-SPT phase corresponding to $H(\lambda_1)$. So we expect a mixed anomaly when there exists no $[\tilde{\omega}]$ such that $[\omega]= [\tilde{\omega}]^2$. Using our experience developed in the previous sections, we can construct a representative $(d+2)$-cocycle $\omega_{d+2}$ of the class in $\mathcal{H}^{d+2}[G\rtimes\mathbb{Z}_2^\mathcal{T},U(1)^*_\pi]$ underlying the mixed anomaly. Here, the notation U$(1)^*_\pi$ means that $\mathbb{Z}_2^\mathcal{T}$ acts on the $G\rtimes \mathbb{Z}_2^\mathcal{T}$ cochains as
\begin{align}
\h\circ\omega_{d+2}((\g_1,\h_1),\dots,(\g_{d+2},\h_{d+2})) = \\
\omega_{d+2}^{\gamma(\h)}((\pi_\h(\g_1),\h_1),\dots,(\pi_\h(\g_{d+2}),\h_{d+2}))
\end{align}
We write the representative $(d+2)$-cocycle $\omega_{d+2}$ of $G\rtimes \mathbb{Z}_2^\mathcal{T}$  as

\begin{align}
\omega_{d+2}((\g_1,\h_1),\dots,(\g_{d+2},\h_{d+2}))=\nonumber \\
\nu_1(\h_1;\g_2,\pi_{\h_2}(\g_3),\pi_{\h_2\h_3}(\g_4),\dots,\pi_{\h_2\dots\h_{d+1}}(\g_{d+2}))
\end{align}
If we take $\nu_1$, with $\h_1$ fixed, to be a $(d+1)$-cocycle of $G$, then the $(d+2)$-cocycle relation for $\omega_{d+2}$ becomes

\begin{align}
1=\h_1\circ\nu(\h_2;\g_3,\pi_{\h_3}(\g_4),\dots,\pi_{\h_3\dots\h_{d+2}}(\g_{d+3})) \nonumber\\
\times \nu^{-1}(\h_1\h_2;\g_3,\pi_{\h_3}(\g_4),\dots,\pi_{\h_3\dots\h_{d+2}}(\g_{d+3}))\nonumber \\
\times \nu(\h_1;\pi_{\h_2}(\g_3),\pi_{\h_2\h_3}(\g_4),\dots,\pi_{\h_2\h_3\dots\h_{d+2}}(\g_{d+3}))
\end{align}
This equation can be identically satisfied if we define $\nu_1$ using the $(d+1)$-cocycle $\omega_{d+1}$ of $G$ as

\begin{align}
\nu_1(\h_1;\g_2,\pi_{\h_2}(\g_3),\dots,\pi_{\h_2\dots\h_{d+1}}(\g_{d+2}) =\nonumber\\
\omega^{(1-\gamma(\h_1))/2}_{d+1}(\pi_{\h_1}(\g_2),\pi_{\h_1\h_2}(\g_3),\dots,\pi_{\h_1\h_2\dots\h_{d+1}}(\g_{d+2}) )
\end{align}
To show that $\omega_{d+2}=\nu_1$ is a trivial cocycle iff $\omega_{d+1}=\tilde{\omega}_{d+1}^2$, with $\tilde{\omega}_{d+1}$ another $(d+1)$-cocycle of $G$, we follow the same steps as in previous section. Assuming $\omega_{d+1}=\tilde{\omega}^2_{d+1}$, we define

\begin{align}
\beta_{d+1}((\g_1,\h_1),\dots,(\g_{d+1},\h_{d+1})) \nonumber \\
= \tilde{\omega}^{-1}_{d+1}(\g_1,\pi_{\h_1}(\g_2),\dots,\pi_{\h_1\dots\h_{d}}(\g_{d+1}))
\end{align}
With this definition, we can write $\nu_1$ as

\begin{align}
\nu_1(\h_1;\g_2,\pi_{\h_2}(\g_3),\dots,\pi_{\h_2\dots\h_{d+1}}\g_{d+2}) = \nonumber \\ 
\h_1\circ \beta_{d+1}((\g_2,\h_2),\dots,(\g_{d+2},\h_{d+2}))
\nonumber \\\times \beta^{(-1)^{d+2}}_{d+1}((\g_1,\h_1),\dots,(\g_{d+1},\h_{d+1}))\nonumber\\
\prod_{q=2}^{d+2}\beta_{d+1}^{(-1)^{q-1}}(\dots,(\g_{q-1}\pi_{\h_{q-1}}(\g_q),\h_{q-1}\h_q),\dots)\, ,
\end{align}
from which it follows that $\nu_1$ is a trivial $(d+2)$-cocycle of $G\rtimes \mathbb{Z}_2^{\mathcal{T}}$. Using the same arguments as in the previous section, we also arrive at the opposite implication that if $\nu_1$ is a trivial $(d+2)$-cocycle, then $\omega_{d+1}=\tilde{\omega}^2_{d+1}$.

\section{Examples} \label{examples}

Above we have argued that there always exists a locality-preserving unitary $\mathcal{U}$ (or anti-unitary $\hat{\mathcal{U}}$) that maps short-range entangled ground states across a quantum critical point between a trivial and non-trivial SPT phase, leading to an additional unitary or anti-unitary $\mathbb{Z}_2$ symmetry at criticality. In general, it is quasi-impossible to explicitely construct this unitary. However, in some special cases the locality-preserving unitary $\mathcal{U}$ or anti-unitary $\hat{\mathcal{U}}$ takes a simple form, as in the examples discussed below. All the models considered here are well-known and have been discussed previously in the literature, but we revisit them with an emphasis on the connections to the general formalism discussed above.

\subsection{Spin-$1/2$ chain}

Let us first consider the classic example of an anti-ferromagnetic spin-$1/2$ chain with Hamiltonian

\begin{equation}\label{spinhalf}
H=\sum_{j} J\vec{S}_{2j}\cdot\vec{S}_{2j+1}+J'\vec{S}_{2j+1}\cdot\vec{S}_{2j+2}\, .
\end{equation}
In general, $J\neq J'$ and this Hamiltonian is only invariant under translation by two sites. We choose our unit cells to be the pairs of spins $(2j,2j+1)$. Note in particular that there is a total integer spin per unit cell, so the internal rotation symmetry group is SO$(3)$. For $J\gg J'$, the ground state will consist of spin singlets within each unit cell, which corresponds to the trivial phase. If $J'\gg J$, on the other hand, the spin singlets form between neighboring unit cells, so the model will be in the Haldane phase.

A translation by one site interchanges $J$ with $J'$, so in this example the locality preserving unitary $\mathcal{U}$ interchanging the trivial with the non-trivial SPT is simply the translation operator. When $J=J'$, the model is the spin-$1/2$ Heisenberg anti-ferromagnet, which is gapless and belongs to the SU$(2)_1$ universality class. At the critical point, translation by one site is a symmetry, and the mixed anomaly between translation symmetry and the on-site spin rotation symmetry was recently shown to underly the Lieb-Schultz-Mattis-Oshikawa-Hastings theorem \cite{LSM,Oshikawa1,Hastings,Furuya,ChengZaletel,Ryu,Jian,Thorngren}.

This mixed anomaly also has a well-known manifestation in the so-called `superspin' description of the SU$(2)_1$ field theory \cite{Senthil}. By writing the SU$(2)$ matrix field $g$ as $g=n^0+i\vec{n}\cdot\vec{\sigma}$, where $\vec{\sigma}$ are the Pauli matrices and $\hat{n}=(n^0,\vec{n})$ is a four-component unit vector field, we can write the SU$(2)_1$ action of the critical point at $J=J'$ as an O$(4)$ non-linear sigma model
\begin{eqnarray}
S & = & S_0 + iS_{WZW}\\
S_0 &= & \frac{\rho_s}{2}\int\mathrm{d}\tau \mathrm{d}x \,\partial_\mu \hat{n}\cdot\partial^\mu \hat{n} \nonumber\\
S_{WZW} & = & \frac{1}{6\pi}\int\mathrm{d}\tau\mathrm{d}x\mathrm{d}u\,\epsilon^{\mu\nu\lambda}\epsilon_{abcd}n^a\partial_\mu n^b\partial_\nu n^c\partial_\lambda n^d \nonumber\, ,
\end{eqnarray}
where as usual, to define the Wess-Zumino-Witten term one extends the vector field $\hat{n}(x,\tau)$ defined with spherical boundary conditions to the solid sphere with coordinates $(x,\tau,u)$ and $u\in[0,1]$, such that $\hat{n}(x,\tau,0)=\hat{n}(x,\tau)$ and $\hat{n}(x,\tau,1)=(1,0,0,0)$. The three-vector $\vec{n}$ has a physical interpretation in terms of the order parameter field of the spin rotation symmetry, while $n^0$ is the VBS order parameter, i.e. the order parameter of translation symmetry. Because of the Wess-Zumino-Witten term, a VBS domain wall binds a spin-$1/2$. The trivial and Haldane phase in Eq. \eqref{spinhalf} are obtained by explicitly breaking translation symmetry, so this spin-$1/2$ is exactly the gapless boundary spin of the Haldane phase.

\subsection{Cluster state}

The Haldane phase is protected even if we break the SO$(3)$ spin rotation symmetry down to a $\mathbb{Z}_2 \times \mathbb{Z}_2$ subgroup consisting of $\pi$ rotations around orthogonal axis. Here, we consider the cluster Hamiltonian \cite{Nielsen}, which is a simpler one-dimensional qubit model that also realizes the SPT phase corresponding to the non-trivial element of $\mathcal{H}^2[\mathbb{Z}_2\times\mathbb{Z}_2,U(1)] = \mathbb{Z}_2$ \cite{SonAmico}. The advantage of the cluster Hamiltonian is that we can explicitly write down a simple finite depth quantum circuit $\mathcal{U}$ mapping it to a trivial SPT Hamiltonian, whereas for the SO$(3)$ spin-$1/2$ chain discussed above $\mathcal{U}$ was the translation operator and could therefore not be witten as a finite depth quantum circuit. The cluster model corresponds to a spin-$1/2$ chain, where we distinguish between the $\sigma$-spins, defined to live on the integer lattice sites, and the $\tau$-spins, which live on half-integer lattice sites. The unit cells of the cluster model, indexed by the integers, therefore contain both a $\sigma$- and a $\tau$-spin. The Hamiltonian takes the following form
 
\begin{equation}
H_{C}=-J\sum_{j} \sigma_j^z\tau^x_{j+1/2}\sigma^z_{j+1}+\tau^z_{j-1/2}\sigma^x_j\tau^z_{j+1/2}
\end{equation}
It has a $\mathbb{Z}_2\times\mathbb{Z}_2$ symmetry generated by $X_1=\prod_j \sigma^x_j$ and $X_2=\prod_j \tau^x_{j-1/2}$. We can now define the quantum circuit $\mathcal{U}$ with depth two as

\begin{equation}
\mathcal{U}=\prod_j CZ_{j,j+1/2}CZ_{j+1/2,j+1}\, ,
\end{equation}
where the matrix $CZ_{j,j'}$ acts on the spins labeled by $j$ and $j'$, and is defined by

\begin{equation}
CZ=|00\rangle\langle00| + |01\rangle\langle01| + |10\rangle\langle10| - |11\rangle\langle11|
\end{equation}
Using $CZ(\sigma^x\otimes\mathds{1})CZ=\sigma^x\otimes\sigma^z$ and $CZ(\mathds{1}\otimes\sigma^x)CZ=\sigma^z\otimes\sigma^x$, one quickly sees that with closed boundary conditions

\begin{equation}
\mathcal{U}^\dagger H_{C}\mathcal{U} = -J \sum_j \tau^x_{j+1/2}+\sigma^x_{j}\, ,
\end{equation}
which is just a trivial paramagnet (see also Ref. \cite{Williamson}). Again with periodic boundary conditions, one also finds that $\mathcal{U}^\dagger X_1\mathcal{U}=X_1$ and $\mathcal{U}^\dagger X_2\mathcal{U}=X_2$. We now define the following Hamiltonian path $H(\lambda)$:

\begin{equation}
H(\lambda) = (1-\lambda)\mathcal{U}^\dagger H_{C}\mathcal{U}+\lambda H_{C}
\end{equation}
Using $\mathcal{U}^2=\mathds{1}$, one sees that this Hamiltonian path satisfies $\mathcal{U}H(\lambda)\mathcal{U}^\dagger = H(1-\lambda)$. It follows that at the transition point $\lambda^*=1/2$, the Hamiltonian has an exact anomalous $\mathbb{Z}_2^3$ symmetry, generated by $X_1$, $X_2$ and $\mathcal{U}$. The transition from the trivial to non-trivial $\mathbb{Z}_2\times\mathbb{Z}_2$-SPT phase was studied in Ref. \cite{Tsui,Tsui2}, and was found to be described by the $c=1$ CFT, similarly to the SO(3) invariant case. 

\subsection{Finite group cohomology fixed point models}

\begin{center}
\begin{figure}
\includegraphics[scale=0.7]{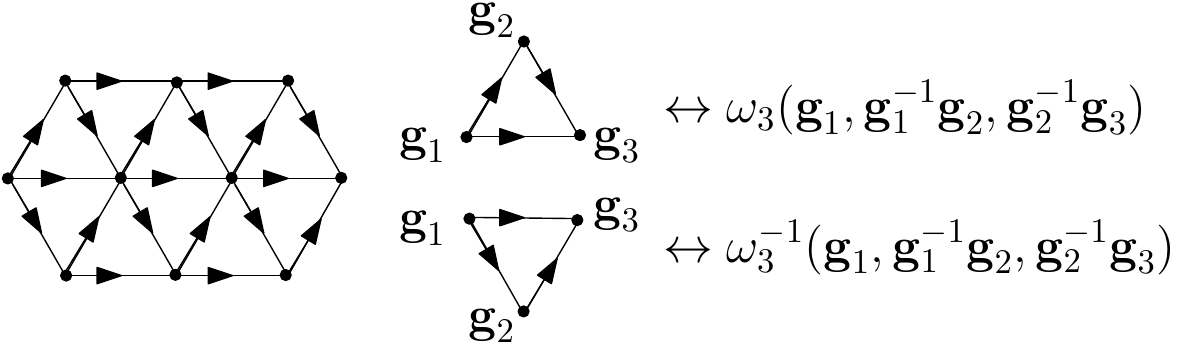}
\caption{The choice of branching structure to define a finite group cohomology fixed point wave function. As explained in the main text, the arrows on the links of the triangular lattice induce a local ordering on the sites around each plaquette. To obtain the non-trivial $G$-SPT ground state wave function one assigns a phase factor $\omega_3$ or $\omega^{-1}_3$ to each plaquette, depending on the choice of branching structure. With our choice, upward pointing triangles get assigned a factor $\omega_3$ and downward pointing triangles a factor $\omega_3^{-1}$.}\label{fig:coh}
\end{figure}
\end{center}

We can generalize the previous discussion of the cluster Hamiltonian to the entire family of cohomology fixed point models for discrete symmetry groups $G$ \cite{Tsui}, introduced in Ref. \cite{ChenWen}. Although this generalization holds independent of the spatial dimension, in this section we restrict to $d=2$ for notational simplicity. In the Hamiltonian formalism, the fixed point models are defined on a triangular lattice, such that on each site there is a Hilbert space with basis states $\{ |\g\rangle, \g\in G\}$. So the local Hilbert space is $|G|$-dimensional, where $|G|$ is the order of the group.  The local symmetry representation is defined to be the left regular representation: $U(\g_1)|\g_2\rangle = |\g_1\g_2\rangle$. The trivial paramagnet ground state is then simply given by

\begin{equation}
|\psi_0\rangle = \left(\frac{1}{\sqrt{|G|}} \sum_{\g\in G} |\g\rangle \right)^{\otimes N}
\end{equation}
The non-trivial $G$-SPT ground state corresponding to $[\omega]\in\mathcal{H}^3[G,U(1)]$, is obtained by acting on $|\psi_0\rangle$ with a quantum circuit $\mathcal{U}$ of depth six. The circuit is defined by picking a representative three-cocycle $\omega_3 \in [\omega]$ and a branching structure on the triangular lattice. The branching structure adopted here is shown on the left of Fig. \ref{fig:coh} by a decoration of the lattice links with arrows. These arrows induce a local ordering on the three sites of each plaquette: the first site has two outgoing arrows, the second has only one outgoing arrow and the third site has none. Using this ordering, the circuit $\mathcal{U}$ is defined by its matrix elements in the group basis as $\mathcal{U}_{\{\g_i,\g'_i\}} = \delta_{\{\g_i\},\{\g'_i\}} \left[\mathcal{U}\right]_{\{\g_i\}}$, where

\begin{align}
\left[\mathcal{U}\right]_{\{\g_i\}}=\prod_{\triangle}\omega_3(\g_1,\g_1^{-1}\g_2,\g_2^{-1}\g_3) \prod_{\bigtriangledown}\omega_3^{-1}(\g_1,\g_1^{-1}\g_2,\g_2^{-1}\g_3)
\end{align}
So $\mathcal{U}$ is diagonal in the group basis, and only changes the phases of the coefficients of $|\psi_0\rangle$ in this basis. In the notation above, $\g_1$ ($\g_2,\g_3$) labels the basis vector on the first (second,third) site in the upward or downward pointing triangle plaquette. Using the three-cocycle relation, one can check that with closed boundary conditions, $\mathcal{U}$ defined in this way commutes with the global symmetry action. We can therefore now write the non-trivial $G$-SPT ground state as

\begin{equation}
|\psi_1\rangle = \mathcal{U}|\psi_0\rangle
\end{equation}
By taking one's favorite Hamiltonian $H_0$ stabilizing the trivial paramagnetic ground state $|\psi_0\rangle$, one can construct the Hamiltonian path

\begin{equation}\label{cohpath}
H(\lambda)=(1-\lambda)H_0 + \lambda \mathcal{U}^\dagger H_0\mathcal{U}\, ,
\end{equation}
which interpolates between the trivial and non-trivial $G$-SPT. Because $\mathcal{U}\mathcal{U}^*=\mathds{1}$ and the left regular representation is real, the Hamiltonian $H(\lambda)$ in Eq. \eqref{cohpath} at the point $\lambda^*=1/2$ has an anti-unitary symmetry $\hat{\mathcal{U}}=\mathcal{U}K$ which commutes with the global symmetry $G$. As discussed above, if the non-trivial $G$-SPT phase $[\omega]$ has no square root, the symmetry group $G\times\mathbb{Z}_2^\mathcal{T}$ at $\lambda=\lambda^*$ has a mixed anomaly. If $[\omega]^2=1$, then $\mathcal{U}=\mathcal{U}^*$, so $H(\lambda^*)$ has a bigger symmetry $G\times\mathbb{Z}_2\times\mathbb{Z}_2^K$, where $\mathbb{Z}_2$ is generated by $\mathcal{U}$, and $\mathbb{Z}_2^K$ by $K$. The symmetry group $G\times\mathbb{Z}_2$ always has a mixed anomaly, irrespective of whether $[\omega]$ has a square root or not. 

Although we can make exact statements about the symmetries and the corresponding anomalies at $\lambda^*$, a downside is that we do not know whether $\lambda^*$ is actually a critical point. An alternative possibility is for example that $\lambda^*$ is a first-order transition point. But for every Hamiltonian path $H(\lambda)$ of the form in Eq. \eqref{cohpath} where a critical point between the trivial and non-trivial SPT phase occurs, this critical point will lie at $\lambda=\lambda^*$ and necessarily come with an exact $\mathbb{Z}_2^{\mathcal{T}}$ symmetry (and also a $\mathbb{Z}_2$ symmetry if $[\omega]^2=1$). This additional symmetry at criticality then leads to possible mixed anomalies with $G$, in line with our general discussion above. 

\subsection{Bosonic integer quantum Hall transition}

\begin{center}
\begin{figure}
\includegraphics[scale=0.7]{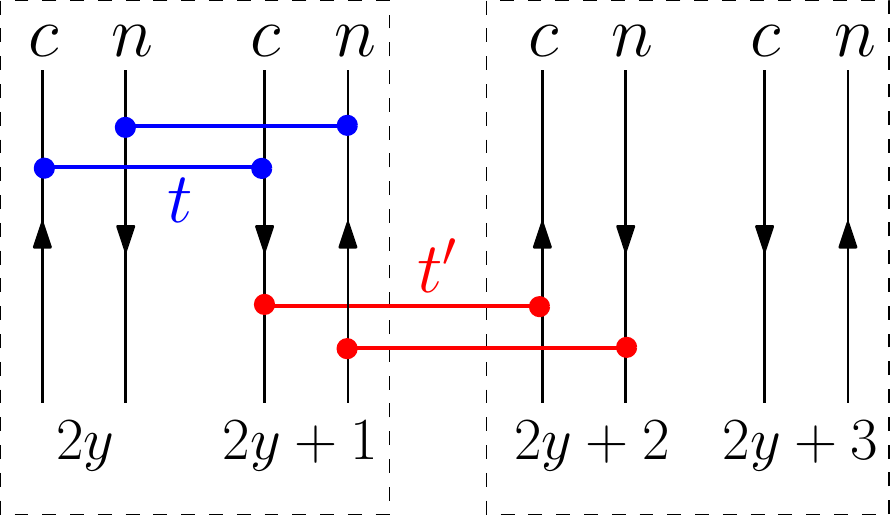}
\caption{Coupled wire construction to study the trivial to BIQH transition. Each wire labeled by $y$ is a non-chiral Luttinger liquid where one chiral boson field is charged under the U$(1)$ symmetry (denoted by $c$), while the field with opposite chirality is neutral ($n$). Each unit cell consists of two wires labeled by $(2y,2y+1)$, and the intra- and inter-unit cell coupling strength between the wires is given by respectively $t$ and $t'$.}\label{fig:wires}
\end{figure}
\end{center}

As a final example, let us consider the well-studied bosonic integer quantum Hall (BIQH) transition \cite{LuVishwanath,ChenWenChiral,SenthilLevinBIQH,VishwanathSenthil,GroverVishwanath,Lu,Geraerdts}. The relevant symmetry group of the BIQH state is simply U$(1)$ charge conservation. With only charge conservation, the group of bosonic U$(1)$-SPT phases in two spatial dimensions is $\mathbb{Z}$, and each SPT phase is uniquely charachterized by its electric Hall conductance $\sigma_{xy}=2n$, $n \in \mathbb{Z}$ \cite{LuVishwanath,ChenWenChiral,SenthilLevinBIQH}. In this section we focus on the transition between the trivial $\sigma_{xy}=0$ phase and $\sigma_{xy}=2$ BIQH phase. 

Because complex conjugation of the charge conservation symmetry acts as $U(\theta)^*=U(-\theta)$, the automorphism $\pi$ is non-trivial here and acts on the angular variable labeling the U$(1)$ group elements as $\pi(\theta)=-\theta$. So we expect a U$(1)\rtimes\mathbb{Z}_2^\mathcal{T}$ symmetry at any critical point between the trivial phase and the BIQH phase with $\sigma_{xy}=2$. This symmetry will have a mixed anomaly because the $\sigma_{xy}=2$ BIQH phase cannot be obtained by stacking two copies of another U$(1)$-SPT. In fact, this already follows from the results of Ref. \cite{VishwanathSenthil}, where it was pointed out that in analogy to the free fermion case, the gapless boundary theory of the 3D bosonic topological insulator is also the critical theory separating the $\sigma_{xy}=0$ and $\sigma_{xy}=2$ BIQH phases.

To investigate the U$(1)\rtimes\mathbb{Z}_2^\mathcal{T}$ symmetry and the associated anomaly in more detail we use a coupled wire approach, as in Refs. \cite{LuVishwanath,VishwanathSenthil,Mross,Mross2,Fuji}. The starting Hamiltonian is of the form $H=\sum_y H_y$, where each $H_y$ describes a one-dimensional (non-chiral) Luttinger liquid at the self-dual radius. We define the unit cells to contain the pairs of wires $(2y,2y+1)$. The U$(1)$ symmetry is taken to act on each wire as it does on the edge of the bosonic integer quantum Hall state, meaning that one chiral boson field has charge one, while the boson field with opposite chirality is charge neutral \cite{LuVishwanath,ChenWenChiral,SenthilLevinBIQH}. For this reason, we denote the chiral boson fields by $\phi_{c,y}$ and $\phi_{n,y}$, where $c$ and $n$ refer to `charged' and `neutral' and $y$ is the wire label. The charge conservation symmetry then acts on the chiral boson fields as

\begin{equation}
\phi_{c,y}\rightarrow \phi_{c,y}+\theta\;,\hspace{0.5 cm} \phi_{n,y}\rightarrow \phi_{n,y}
\end{equation}
Importantly, we have to ensure that within one unit cell the U$(1)$ symmetry can be implemented locally by taking $\phi_{c,2y}$ and $\phi_{c,2y+1}$ to have opposite chirality. Having fixed the notation, we can now write down the Hamiltonian of the wire labeled by $y$ as

\begin{equation}
H_y = \frac{1}{2\pi}\int\mathrm{d}x\;(\partial_x\phi_{c,y})^2 +(\partial_x\phi_{n,y})^2\, ,
\end{equation}
where the chiral boson fields $\phi_{c,y}$ and $\phi_{n,y}$ satisfy the commutation relations

\begin{eqnarray}\label{comm}
[\phi_{c,y}(x),\phi_{c,y'}(x')]  & = & (-1)^{y}\delta_{yy'}i\frac{\pi}{2}\,\text{sgn}(x-x')\\
 \left[\phi_{n,y}(x),\phi_{n,y'}(x')\right] & =  & (-1)^{y+1}\delta_{yy'}i\frac{\pi}{2}\,\text{sgn}(x-x')  \\
\left[\phi_{c,y}(x),\phi_{n,y'}(x')\right] & = & 0
\end{eqnarray}
We now couple the wires with the following U$(1)$-symmetry respecting boson hopping:

\begin{align}
-\sum_y \Big[ t\left(e^{-i\phi_{c,2y}}e^{i\phi_{c,2y+1}}+e^{-i\phi_{n,2y}}e^{i\phi_{n,2y+1}}\right)+ \nonumber \\
t'\left(e^{-i\phi_{c,2y+1}}e^{i\phi_{c,2y+2}}+e^{-i\phi_{n,2y+1}}e^{i\phi_{n,2y+2}}\right) + h.c.\Big]
\end{align}
So $t$ is the coupling strength between wires within the same unit cell, while $t'$ is the coupling strength between wires in neighboring unit cells. See also Fig. \ref{fig:wires} for a schematic representation of the coupled wire construction. For $t\gg t'$, the intra-unit cell backscattering dominates and the system will form a trivial insulator. For $t'\gg t$ on the other hand, the backscattering between wires in neighboring cells dominates. In analogy to the spin-$1/2$ chain discussed above, this will lead to the non-trivial $\sigma_{xy}=2$ SPT. This can be understood from the fact that with open boundaries, there will be one unpaired Luttinger liquid along the boundary, which is the gapless edge mode of the BIQH phase \cite{VishwanathSenthil}. 

When $t\neq t'$, the coupled wire system is clearly not invariant under a translation by one site (or wire). As can be seen from the commutation relations in Eq. \eqref{comm}, the following anti-unitary operation interchanges $t$ with $t'$:

\begin{equation}\label{QHsymm}
\phi_{c/n,y} \rightarrow -\phi_{c/n,y+1}\;,\hspace{0.5 cm} i\rightarrow -i
\end{equation}
It therefore maps the trivial phase to the BIQH phase and vice versa.  So it is implemented by the locality preserving anti-unitary operator $\hat{\mathcal{U}}$ introduced in the previous sections. It follows that the transition between the trivial phase and the BIQH phase occurs at $t=t'$, at which point the anti-unitary map in Eq. \eqref{QHsymm} becomes a symmetry.

To see how the $\mathbb{Z}_2^\mathcal{T}$ is implemented on the level of the field theory describing the transition, we first go to the SU$(2)_1$ description of the self-dual compact boson. So let us define the non-chiral boson fields

\begin{eqnarray}
\varphi_y(x) & = & \phi_{c,y}+\phi_{n,y} \\
\theta_y(x) & = & \phi_{c,y}-\phi_{n,y}\, ,
\end{eqnarray}
and use these to construct the following SU$(2)$ matrix:

\begin{equation}
U_y = \frac{1}{\sqrt{2}}\left(\begin{matrix}e^{i\varphi_y} & e^{i\theta_y}\\- e^{-i\theta_y} & e^{-i\varphi_y} \end{matrix}\right)
\end{equation}
The SU$(2)_1$ action for each wire can then be written as

\begin{eqnarray}
S_y & = & S_{y,0} + iS_{y,WZW} \\
S_{y,0} & = & \frac{\rho_s}{2}\int\mathrm{d}\tau \mathrm{d}x\,\text{tr}\left(\partial_\mu U_y^\dagger \partial^\mu U_y\right)\\
S_{y,WZW} & = & \frac{1}{12\pi}\int\mathrm{d}\tau \mathrm{d}x\mathrm{d}u\,\epsilon^{\mu\nu\lambda}\text{tr}\left(U_y^\dagger\partial_\mu U_yU_y^\dagger\partial_\nu U_yU_y^\dagger\partial_\lambda U_y\right)\nonumber\, ,
\end{eqnarray}
where as before, the Wess-Zumino-Witten term is defined by extending the matrix field $U_y$ onto a solid sphere with additional coordinate $u$. The problem of obtaining a $(2+1)$-d continuum field theory for the coupled SU$(2)_1$ wires with $t=t'$ was solved in Ref. \cite{Senthil}. The slowly varying field describing the long-wavelength modes in the $(2+1)$-d system is obtained by defining $g_y$ as 

\begin{equation}\label{g}
g_y = \begin{cases} U_y, & y=2n\\ 
  U^\dagger_y, &  y=2n+1 \end{cases}
\end{equation}
and then promoting $y$ to a continuum variable to arrive at the SU$(2)$ matrix field $g(\tau,x,y)=g_y(\tau,x)$. By again writing the matrix field in terms of a four-component unit vector field $\hat{n}$ as $g=n^0 + i\vec{n}\cdot\vec{\sigma}$, the authors of Ref. \cite{Senthil} arrived at the following $(2+1)$-d O$(4)$ non-linear sigma model 

\begin{eqnarray}
S & = & S_0 + i\Theta S_1 \\
S_0 & = & \frac{\rho_s}{2}\int\mathrm{d}\tau\mathrm{d}^2x\,\partial_\mu \hat{n}\cdot  \partial^\mu \hat{n}\\
S_1 & = &\frac{1}{12\pi^2} \int\mathrm{d}\tau\mathrm{d}^2x\,\epsilon^{\mu\nu\lambda}\epsilon_{abcd}n^a\partial_\mu n^b \partial_\nu n^c \partial_\lambda n^d\, ,
\end{eqnarray}
with $\Theta = \pi$. The $\mathbb{Z}_2^\mathcal{T}$ symmetry \eqref{QHsymm} acts on $U_y$ as $U_y\rightarrow U_{y+1}$. So from Eq. \eqref{g}, we see that it acts on the $(2+1)$-d continuum fields as $g\rightarrow g^\dagger$ and $(n^0,n^1,n^2,n^3)\rightarrow (n^0,-n^1,-n^2,-n^3)$.

The O$(4)$ sigma model with a theta term at $\Theta=\pi$ has a dual description in terms of $N_f=2$ QED$_3$ \cite{AliceaMotrunich,Senthil,Seiberg,WangNahum}, which has been proposed to describe the BIQH transition in Refs. \cite{Lu,GroverVishwanath}. To obtain this fermionic description in the coupled wire formalism, we follow Refs. \cite{Mross,Mross2,BosonicCFL} in their derivation of the fermionic particle-vortex duality \cite{WangSenthilX,MetlitskiVishwanath,Karch,Son}, and define following dual boson field $\tilde{\phi}_{c,y}$

\begin{equation}\label{duality}
\tilde{\phi}_{c,y} = \sum_{y'\neq y}\text{sgn}(y-y')(-1)^{y'}\phi_{c,y'}
\end{equation}
This dual boson field is also chiral, but with opposite chirality, as can be seen from its commutation relations

\begin{equation}
[\tilde{\phi}_{c,y}(x),\tilde{\phi}_{c,y'}(x')] = - [\phi_{c,y}(x),\phi_{c,y'}(x')]
\end{equation}
Interestingly, the dual field $\tilde{\phi}_{c,y}$ is neutral under the global U$(1)$ symmetry. By applying the linear transformation

\begin{eqnarray}\label{phipm}
\phi_{\pm,y} & = & \tilde{\phi}_{c,y}\pm \phi_{n,y}
\end{eqnarray}
we can now define the chiral fermion fields

\begin{eqnarray}
\psi_{\pm,y}(x)=\eta_{\pm,y} e^{i\phi_{\pm,y}(x)}\, ,
\end{eqnarray}
where $\eta_{\pm,y}$ are Klein factors to ensure that the different fermion species, and also fermion fields on different wires, anti-commute. Because of the duality transformation in Eq. \eqref{duality}, the Lagrangian of the coupled-wire system becomes non-local in terms of the fermion fields. However, it was shown by the authors of Ref. \cite{Mross} that one can recover locality by introducing a gauge field $a_\mu$. By defining the two-component fermions $\Psi_{\pm}(x,y) = \left(\psi_{\pm,2y}(x)\, ,\;\psi_{\pm,2y+1}(x)\right)^T$ and again promoting $y$ to a continuum variable, one finds the non-compact $N_f=2$ QED$_3$ continuum theory \cite{Mross,BosonicCFL,GroverVishwanath,Lu}

\begin{eqnarray}\label{eq:QED}
\mathcal{L} & = & \sum_{\tau=\pm}\bar{\Psi}_{\tau}\gamma^\mu \left(-i\partial_\mu+a_\mu\right)\Psi_{\tau} \\ & & +\frac{1}{2\pi}\epsilon^{\mu\nu\lambda}A_\mu\partial_\nu a_\lambda - \frac{1}{4\pi}\epsilon^{\mu\nu\lambda}A_\mu\partial_\nu A_\lambda\, \nonumber
\end{eqnarray}
where $\bar{\Psi}=\Psi^\dagger\gamma^0$ and the $2\times 2$ Dirac matrices are given by $\{\gamma^0,\gamma^1,\gamma^2\}=\{\sigma^x,-i\sigma^y, i\sigma^z \}$. We have also added a probe gauge field $A_\mu$ for the global U$(1)$ symmetry, to emphasize that the fermion fields are charge neutral and that the global symmetry charge is carried by the vortices of the dynamical gauge field $a_\mu$ \cite{WangSenthilX,MetlitskiVishwanath,Karch,Son}. 

The $\mathbb{Z}_2^\mathcal{T}$ symmetry \eqref{QHsymm} acts on the dual chiral boson field as $\tilde{\phi}_{c,y}\rightarrow \tilde{\phi}_{c,y+1}$, from which its action on the fields $\phi_{\pm,y}$ follows:

\begin{equation}
\phi_{\pm,y}\rightarrow \phi_{\mp,y+1}\,,\hspace{0.5 cm} i\rightarrow -i
\end{equation}
So we see that $\mathbb{Z}_2^\mathcal{T}$ acts as a particle-hole symmetry on the fermion fields which also interchanges the two fermion flavors.  Defining $\Psi = \left(\Psi_+\, ,\;\Psi_-\right)^T$, we can therefore write the $\mathbb{Z}_2^\mathcal{T}$ action as
\begin{equation}\label{eq:symmact}
\Psi\rightarrow \tau^x\sigma^x\Psi^\dagger\, ,\hspace{0.5 cm} i\rightarrow -i
\end{equation}
From the Lagrangian in Eq. \eqref{eq:QED} we see that $\mathbb{Z}_2^\mathcal{T}$ is a $\mathcal{CT}$ symmetry, where $\mathcal{C}$ is gauge-charge conjugation and $\mathcal{T}$ is time-reversal. 

Our choice of applying the duality transformation \eqref{duality} to $\phi_{c,y}$ was arbitrary, and we could instead have applied it to the charge neutral boson fields $\phi_{n,y}$. From Eq. \eqref{phipm}, we see that in this case the fermions $\Psi_{\pm}$ will be charged under the global U$(1)$ symmetry, but the two flavors will have opposite gauge charge. After a particle-hole symmetry on one of the flavors, we recover $N_f=2$ QED$_3$, but with the two fermion flavors having opposite global U$(1)$ charge. This is the self-duality of $N_f=2$ QED$_3$ \cite{XuYou,ChengXu,Mross,Mross2,Hsin,Seiberg,WangNahum}. The $\mathbb{Z}_2^\mathcal{T}$ symmetry acts in the same way on the dual fermions as in Eq. \eqref{eq:symmact}, which is consistent with the global U$(\theta) = e^{i\theta\tau^z}$ symmetry action as this indeed forms a representation of U$(1)\rtimes\mathbb{Z}_2^\mathcal{T}$.  

Before turning to the final section, we note that the $\Theta=\pi$ O$(4)$ sigma model and $N_f=2$ QED$_3$ have been discussed in great detail in the context of deconfined quantum criticality in Refs. \cite{WangNahum,Dualityreview}. It was also pointed out that these field theories describe a critical point between bosonic SPT phases, which can be obtained by adding a relevant operator that breaks an anomalous symmetry \cite{WangNahum,Dualityreview}.

\section{Discussion}\label{discussion}

We have argued that at critical points between a trivial and non-trivial bosonic $G$-SPT phase there is a minimal symmetry group $G\times\mathbb{Z}_2$, $G\times\mathbb{Z}_2^\mathcal{T}$ or $G\rtimes\mathbb{Z}_2^\mathcal{T}$, depending on whether the $G$-SPT phase squares to the trivial phase, and the type of local symmetry representation. For bosonic SPT phases described by group cohomology, we have identified when the symmetry group at criticality has a mixed global symmetry anomaly. When such a mixed symmetry anomaly is present, this provides a non-trivial consistency check on any field theory which in the IR is supposed to flow to the CFT describing the critical point.

\begin{center}
\begin{figure}
\includegraphics[scale=0.6]{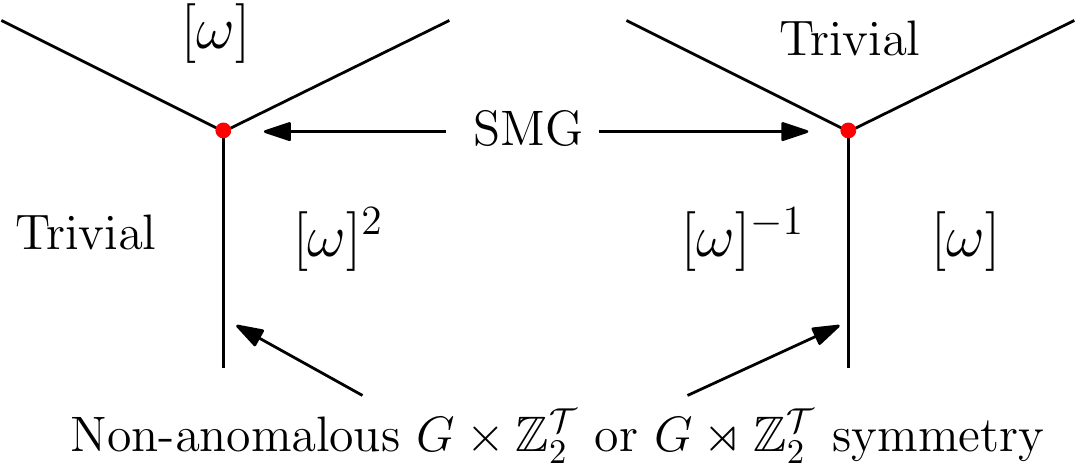}
\caption{An illustration of the symmetric mass generation (SMG) transition. The multi-critical point depicted by a red circle can be interpreted as a SMG transition because the trivial-$[\omega]^2$ critical line is anomaly-free. By stacking the phase diagram with the $G$-SPT $[\omega]^{-1}$, the anomaly-free critical line separates the $G$-SPT phases $[\omega]^{-1}$ and $[\omega]$. This is the SMG scenario discussed in Ref. \cite{You}.}\label{fig:SMG}
\end{figure}
\end{center}

If there is no mixed anomaly, then an analog of the symmetric mass generation (SMG) transition is possible \cite{BenTov,Catterall,Venkitesh,Schaich,Ayyar,HeWu,Huffman,YouHeXu,You}. Consider a two-parameter phase diagram where an anomaly-free critical line between two bosonic SPT phases ends at a multi-critical point, and assume that all the gapped phases surrounding this multi-critical point are all short-range entangled. See Fig. \ref{fig:SMG} for an illustration of this scenario. In this case, the multi-critical point can be interpreted as a SMG point, where there is a transition from the anomaly-free critical theory to a short-range entangled symmetric phase. In Ref. \cite{You}, it was shown that this scenario occurs for the critical theory separating two bosonic SO$(4)$-SPT phases which are each others inverse. In that case, the phase transition between the SPT phases is described by eight Dirac cones, which can generate a mass without breaking any symmetries. The transition from the semi-metal to the featureless symmetric phase was proposed to be described by a SU$(4)$ QCD-Higgs theory \cite{YouHeXu,You}. To see how this relates to our findings, recall that we showed that the critical theory separating the trivial phase and the $G$-SPT $[\omega]^2$ has no mixed anomaly between SO$(4)$ and the emergent anti-unitary $\mathbb{Z}_2^\mathcal{T}$ symmetry \footnote{If $[\omega]^4= 1$, there would be a mixed anomaly between SO$(4)$ and an emergent unitary $\mathbb{Z}_2$ symmetry, but this is not the case for SO$(4)$ SPTs.}. So let us consider the scenario where the anomaly-free critical line ends at a multi-critical point, where we enter the $G$-SPT phase $[\omega]$, which is exactly the SO$(4)$-SPT phase corresponding to the fixed point of the $\hat{\mathcal{U}}$ operator. This is shown on the left of Fig. \ref{fig:SMG}. As mentioned in the introduction, we can stack a SO$(4)$-SPT phase $[\omega]^{-1}$ on top of the entire phase diagram without changing the critical fluctuations. Now the anomaly-free critical line separates the phases labeled by $[\omega]^{-1}$ and $[\omega]$, and ends in the trivial phase. See the right-hand side of Fig. \ref{fig:SMG} for an illustration. This is exactly the type of phase diagram with SMG explored in Ref. \cite{You}.

An important open question is how to generalize our results to interacting fermion SPT phases. We expect that at least part of our reasoning can also be applied to fermionic SPT phases, because just as in the bosonic case they only have short-range entanglement. To motivate this expectation, see in particular Refs. \cite{FidkowskiPo} and \cite{Ellison}. In Ref. \cite{FidkowskiPo}, one-dimensional fermionic locality-presering unitaries are studied. A particularly interesting example of such a fermionic locality-preserving unitary is the `Majorana translation operator', which maps a trivial one-dimensional fermion chain to the Majorana chain and vice versa. In Ref. \cite{Ellison}, an explicit procedure was given to disentangle supercohomology \cite{Supercohomology} SPT ground states in two spatial dimensions. The corresponding `disentangling operator' was found to be a quantum circuit, and is therefore another example of a fermionic locality-preserving unitary that maps a trivial state to an SPT ground state.

One potential problem for a naive generalization of our approach is that complex conjugation no longer acts as the inverse on the group of fermionic SPT phases. Also, fermionic systems can be subject to different types of symmetry anomalies which do not occur in bosonic systems, see for example Refs. \cite{ChengLSM,FidkowskiMetlitski,Jones,Sullivan}. Based on our experience with bosonic SPT phases, we expect that also in the fermionic case the recently constructed exactly solvable fixed-point Hamiltonians  \cite{Tarantino,Ware,WangNing,Nathanan} can be of great use for gaining insights on how to proceed in this matter.

\emph{Acknowledgements --} The author would like to thank Paul Fendley, Jutho Haegeman, Masaki Oshikawa, Thomas Scaffidi, Laurens Vanderstraeten and Frank Verstraete for helpful discussions, and especially Michael Zaletel for his comments on a previous version of this manuscript. NB was supported by the DOE, office of Basic Energy Sciences under contract no. DEAC02-05-CH11231.

\appendix

\section{Group cohomology review}\label{AppCoh}

Given a discrete symmetry group $G$ and an abelian group $M$, we define $n$-cochains $\omega_n$ to be maps from $G^{\times n}$ to $M$. The set of $n$-cochains, denoted by $C^n(G,M)$, forms an abelian group with multiplication

\begin{equation}
(\omega_n\cdot\omega'_n)(\g_1,\dots,\g_n) = \omega_n(\g_1,\dots,\g_n)\omega'_n(\g_1,\dots,\g_n)
\end{equation}
and identity $\omega_n\equiv 1$, where $1$ is the identity group element of $M$. Given an action of $G$ on $C^n(G,M)$, denoted as $\g\circ \omega_n$, we can now define a coboundary operator $\mathrm{d}:C^n(G,M)\rightarrow C^{n+1}(G,M)$ as follows

\begin{align}
\mathrm{d}\omega_n(\g_1,\dots,\g_{n+1}) = \g_1\circ \omega_n(\g_2,\dots,\g_{n+1}) \\
\prod_{j=1}^{n}\omega_n^{(-1)^j}(\g_1,\dots,\g_{j-1},\g_{j}\g_{j+1},\dots,\g_{n+1}) \nonumber\\
\omega_n^{(-1)^{n+1}}(\g_1,\dots,\g_n)
\end{align}
Provided that $\g_1\circ (\g_2\circ \omega_n)=(\g_1\g_2)\circ \omega_n$, the coboundary map has the important property that $\mathrm{d}\mathrm{d}\omega_n\equiv 1$. We define a $n$-cocycle to be an element $\omega_n$ of $C^n(G,M)$ satisfying $\mathrm{d}\omega_n=1$. The set of $n$-cocycles is denoted as $\mathcal{Z}^n(G,M_\circ)$, where the notation $M_\circ$ reminds us of the action of $G$ on the cochains. 

If a $n$-cochain $\omega_n$ can be written as $\omega_n=\mathrm{d}\omega_{n-1}$, we say that $\omega_n$ is a $n$-coboundary. Because $\mathrm{d}\mathrm{d}\omega_n=1$, every $n$-coboundary is also a $n$-cocycle. The set of $n$-coboundaries is denoted as $\mathcal{B}^n(G,M_\circ)$. 

If we now require that $(\g\circ\omega_n)\cdot(\g\circ\omega'_n)=\g\circ(\omega_n\cdot\omega'_n)$, it holds that $\mathrm{d}\omega_n\cdot \mathrm{d}\omega'_n= \mathrm{d}(\omega_n\cdot\omega'_n)$, such that both $\mathcal{Z}^n(G,M_\circ)$ and $\mathcal{B}^n(G,M_\circ)$ form abelian groups. One can easily recognize $\mathcal{B}^n(G,M_\circ)$ to be a normal subgroup of $\mathcal{Z}^n(G,M_\circ)$, and so we define the $n$-th cohomology group as

\begin{equation}
\mathcal{H}^n[G,M_\circ] = \mathcal{Z}^n(G,M_\circ)/\mathcal{B}^n(G,M_\circ)
\end{equation}
This means that the cohomology group labels equivalence classes of $n$-cocyles that differ by a $n$-coboundary. The algebraic definition of group cohomology can be generalized to continuous groups, by imposing continuity conditions on the cocycles.

\section{Dimensional reduction procedure}

In this appendix, we use a generalization of the dimensional reduction procedure presented in Ref. \cite{Else}, to show that $\mathcal{H}^1[\mathbb{Z}_2,\mathcal{H}^2[G,U(1)]]$ indeed describes the mixed anomaly at a critical point between the trivial phase in one spatial dimension and a non-trivial $G$-SPT phase which squares to the trivial phase. 

We first define for $\h \in H=\mathbb{Z}_2$

\begin{equation}
\mathcal{U}^\h=\begin{cases}\mathds{1}\,, & \h \text{ is the identity group element} \\ \mathcal{U}\,, & \h \text{ is the non-trivial group element} \end{cases}\, ,
\end{equation}
where as in the main text, $\mathcal{U}$ is the locality-preserving unitary mapping the trivial SPT to the non-trivial SPT. The symmetry group at criticality is $G\times \mathbb{Z}_2$, and a general group element is represented as $(\g,\h)=\hat{U}(\g)\mathcal{U}^\h$, where $\hat{U}(\g)$ is the tensor product of the local symmetry action $U(\g)$ over all the lattice sites. Let us work on an infinite chain, and define $\hat{U}(\g)\big|_L$ to be the symmetry action on the half-infinite chain left of the lattice site at position $x=0$. Because $\mathcal{U}$ is locality-preserving and $\mathcal{U}$ commutes with $\hat{U}(\g)$, we expect that

\begin{equation}\label{comm}
\mathcal{U}^{\h\dagger}\hat{U}(\g)\big|_L\mathcal{U}^\h = \hat{U}(\g)\big|_L V_\h(\g)\,,
\end{equation}
where $V_\h(\g)$ is a unitary operator with support on a finite region around $x=0$. We have no proof that this is the most general possibility, but we will assume this here. Clearly, $V_\h(\g)\equiv \mathds{1}$ when $\h$ is the identity group element. 

The left-hand side of Eq. \eqref{comm} also forms a representation of $G$. Because we are working on a half-infinite chain, the right-hand side of Eq. \eqref{comm} can satisfy the group multiplication rule up to a phase (a more rigorous way to approach this would be to start with a restriction of $\hat{U}(\g)$ to a finite interval and then take the length of the interval to infinity). This phase ambiguity is important here. Because $\mathcal{U}$ maps a trivial state to a $G$-SPT ground state, $V_\h(\g)$ will be a non-trivial projective representation of $G$ when $\h$ is the non-trivial group element of $\mathbb{Z}_2$. 

We can also interpret Eq. \eqref{comm} as saying that with $\hat{U}(\g)$ restricted to a half-infinite chain, the $G\times H$ multiplication rules are satisfied up to a boundary operator in the following way

\begin{align}\label{boundop}
\left( \hat{U}(\g_1)\big|_L\mathcal{U}^{\h_1}\right)\left(\hat{U}(\g_2)\big|_L\mathcal{U}^{\h_2}\right) = \\ \left( \hat{U}(\g_1\g_2)\big|_L\mathcal{U}^{\h_1}\mathcal{U}^{\h_2}\right)\Omega((\g_1,\h_1),(\g_2,\h_2))\, ,\nonumber
\end{align}
where it follows from Eq. \eqref{comm} that $\Omega$ takes the form

\begin{equation}
\Omega((\g_1,\h_1),(\g_2,\h_2)) = \mathcal{U}^{\h_2\dagger} V^\dagger_{\h_1}(\g_2)\mathcal{U}^{\h_2}
\end{equation}
Because $\mathcal{U}^{\h_2}$ is locality preserving, $\Omega$ is indeed a boundary operator supported only on a finite region around $x=0$. Equation \eqref{boundop} is exactly the starting point of the dimensional reduction procedure of Ref. \cite{Else}, but the difference between our approach and that of Ref. \cite{Else} is that here we do not restrict $\mathcal{U}^\h$. We can nevertheless proceed in analogy to Ref. \cite{Else}, and study the associativity properties of the restricted symmetry operators. To this end, let us consider

\begin{equation}\label{eq:asso}
\left(\hat{U}(\g_1)\big|_L \mathcal{U}^{\h_1}\right)\left(\hat{U}(\g_2)\big|_L \mathcal{U}^{\h_2}\right)\left(\hat{U}(\g_3)\big|_L \mathcal{U}^{\h_3}\right)\, ,
\end{equation}
and evaluate the product in two different ways. In the first way of evaluating the product \eqref{eq:asso}, one finds

\begin{equation}\label{ev1}
\hat{U}(\g_1\g_2\g_3)\big|_L \mathcal{U}^{\h_1}V^\dagger_{\h_1}(\g_3)V^\dagger_{\h_1}(\g_2)\mathcal{U}^{\h_2}V^\dagger_{\h_2}(\g_3)\mathcal{U}^{\h_3}
\end{equation}
The second way of evaluating \eqref{eq:asso} gives

\begin{equation}\label{ev2}
\hat{U}(\g_1\g_2\g_3)\big|_L \mathcal{U}^{\h_1}V^\dagger_{\h_1}(\g_2\g_3)\mathcal{U}^{\h_2}V^\dagger_{\h_2}(\g_3)\mathcal{U}^{\h_3}
\end{equation}
Now because $V_{\h_1}(\g)$ is a projective representation when $\h_1$ is the non-trivial group element of $\mathbb{Z}_2$, we find that Eq. \eqref{ev1} is equal to Eq. \eqref{ev2} up to the phase factor $\omega_3((\g_1,\h_1),(\g_2,\h_2),(\g_3,\h_3)) = \nu_1(\h_1,\g_2,\g_3)$, where $\nu_1(\h_1,\g_2,\g_3)$ with fixed non-trivial $\h_1$ is a representative $2$-cocycle of the non-trivial $G$-SPT, and $\nu_1(\h_1,\g_2,\g_3)=1$ when $\h_1$ is the identity group element. 

Note that at up to now we did not make use of the fact that the non-trivial $G$-SPT phase squares to the trivial phase. And even with the $G$-SPT phase squaring to the trivial phase, it is important to remember that $\mathcal{U}^\h$ does not actually form a representation of $\mathbb{Z}_2$, i.e. in general $\mathcal{U}^{\h_1}\mathcal{U}^{\h_2}\neq \mathcal{U}^{\h_1\h_2}$. But as we just learned, this does not stop us from defining the phase factor $\nu_1(\h_1,\g_2,\g_3)$. However, in order to show that $\nu_1$ is a $3$-cocycle of $G\times\mathbb{Z}_2$, the fact that the $G$-SPT phase squares to the trivial phase is crucial. One can attempt to repeat the proof of Ref. \cite{Else} in order to show that $\nu_1$ satisfies the $3$-cocycle relation, but this will produce the result that $\nu_1$ is a $3$-cocycle of $G\times \mathbb{Z}$, because $\mathcal{U}^\h$ is not a representation of $\mathbb{Z}_2$. In order to arrive at a $3$-cocycle of $G\times\mathbb{Z}_2$, one first iterates Eq. \eqref{comm} twice to obtain

\begin{align}
\mathcal{U}^{\h_2\dagger}\mathcal{U}^{\h_1\dagger}\hat{U}(\g)\big|_L\mathcal{U}^{\h_1} \mathcal{U}^{\h_2}  = \nonumber\\
\hat{U}(\g)\big|_L \left[ V_{\h_2}(\g)\mathcal{U}^{\h_2\dagger}V_{\h_1}(\g)\mathcal{U}^{\h_2}\right]
\end{align}
Using the same logic as before, and the fact that the non-trivial $G$-SPT phase squares to the trivial phase, the operator between brackets in the second line will be a linear (projective) representation of $G$ when $\h_1\h_2$ is the identity (non-trivial) group element of $\mathbb{Z}_2$. This fact has to be used to show that there exists a $2$-coboundary transformation (the same discussed in the main text) that turns $\nu_1$ into a $3$-cocycle of $G\times\mathbb{Z}_2$.

\bibliography{bibliography}

\end{document}